\documentclass[a4paper,10pt]{article}
\usepackage{authblk}
\usepackage{amssymb} 
\usepackage{amsmath} 
\usepackage{graphicx}

\title{Semi-Markovian Dynamics of a Self-Propelled Particle in a Confined Environment: A Large-Deviation Study} 
\author{Shabnam Sohrabi and  Farhad H. Jafarpour\footnote{f.jafarpour@basu.ac.ir}}
\affil{Faculty of Science, Bu-Ali Sina University ,Hamedan, Iran} 

\begin{document}   
\maketitle
\begin{abstract}
We study the large deviations of the time-integrated current for a self-propelled particle moving within a
confined environment. The dynamics is modeled as a semi-Markovian process, where the transitions between a 
\textit{normal running phase} (Phase $0$) and a \textit{wall-attached phase} (Phase $1$) are governed by time-dependent 
reset probabilities. We study two different examples: In the first case, the particle undergoes a biased random walk in 
Phase $0$, while it intermittently resets and interacts with the container boundaries, remaining stationary in Phase $1$.
In this scenario, the reset probabilities for transitions between the two phases follow an ``aging'' logic.
In the second case, the particle alternates between two active phases: a Markovian Phase $0$ characterized 
by memoryless, downstream-biased motion, and a semi-Markovian Phase $1$ with a reversed, upstream bias representing
boundary-attached navigation. Here, we assume a time-independent survival probability in Phase $0$ and a time-dependent one 
in Phase $1$. By analyzing the Scaled Cumulant Generating Function (SCGF) in the long-time limit, we derive 
the conditions for Dynamical Phase Transition (DPT)s in the fluctuations of the particle velocity. We demonstrate that, depending on the 
aging strength, the system exhibits either discontinuous (first-order) or continuous (second-order) DPTs. 
Analytical predictions are validated via computer simulations.
\end{abstract}
\section{Introduction}\label{sec:1} 
The investigation of self-propelled particle dynamics, such as those exhibited by bacteria, in confined environments has
received significant attention over the past decade \cite{Du2021}.  Experimental observations on mammalian sperm cells show that individuals
exhibit random downstream drift in the channel center but reorient and swim upstream near walls,
where their alignment becomes increasingly stable over time due to hydrodynamic torques and surface
interactions \cite{Ka2014}. Analogous upstream persistence has
been documented in bacterial rheotaxis, involving both bulk mechanisms in helical swimmers \cite{Ma2012} 
and surface-enhanced mechanisms where proximity to boundaries suppresses reorientation
events \cite{Mo2014}. In the latter, surface trapping fosters prolonged retention, with
studies on rheotactic invasion revealing that these interactions facilitate long upstream runs
characterized by heavy-tailed or power-law distributions of run times \cite{Fi2020}. Furthermore, state-dependent 
bias switches coupled with asymmetric memory structures in
simple 1D random walk models provide a powerful framework for reproducing upstream accumulation and
counterflow invasion, highlighting the role of non-Markovian persistence in rheotactic navigation \cite{Fi2020}.

In this paper, we introduce a minimal one-dimensional model  in discrete time and discrete space for a self-propelled 
particle in a confined environment that alternates between a normal running phase (Phase $0$) and a wall-attached phase (Phase $1$) 
via stochastic resets. We aim to investigate how time-dependent reset protocols influence the fluctuations of time-integrated 
observables, such as the particle's displacement \cite{To2009,To2025}. Unlike previous studies of similar models that considered constant 
reset rates \cite{Ev2020}, we focus on the role of aging. By incorporating time-dependent reset probabilities, we describe 
the system through semi-Markovian dynamics and analyze the resulting statistical properties. 

We analyze two distinct examples in detail: in the first case, the particle performs a biased random walk in Phase $0$. 
Phase $1$ represents a wall-attached state in which the particle is immobilized at the container boundary. We assume 
that the residence time in each phase is governed by a heterogeneous distribution, meaning that the probability of 
remaining in a phase or transitioning to the other phase depends on the time already spent in the current state. 
Specifically, we assume that the particle possesses an internal clock that resets itself upon every reset. The time-dependence of the reset 
probability is inspired by \cite{Ko2018}, where the authors investigated semi-Markovian intracellular transport involving sub-diffusion 
and run-length-dependent detachment rates. The semi-Markovian nature of the process arises because transition rates 
depend on the residence time within the current phase rather than the total elapsed time. We assume that every excursion 
in Phase $0$ or Phase $1$ ends with a reset to the other phase. It turns out that both first-order and second-order
DPT can be observed in the fluctuations of displacement of the particle depending on the aging strength. 
Furthermore, we find that the Gallavotti-Cohen symmetry holds in this case. In the second case, we propose a  minimal model 
of rheotaxis where the particle alternates between two active phases: a Markovian Phase $0$ characterized by memoryless, downstream-biased
motion in the bulk, and a semi-Markovian Phase $1$ with a reversed, upstream bias representing
boundary-proximity navigation. As in the previous case, we assume that every excursion in each phase ends with 
a rest to the other phase. This framework generalizes an ordinary time-independent reset model
to a compound process where asymmetric memory structures compete alongside opposing directional
biases. The physical basis of this model is found in the experimental upstream rheotaxis of
microswimmers; while bulk transport is stochastic and memoryless, contact with surfaces triggers
reorientation against the flow. Crucially, empirical statistics of surface residence times often
exhibit heavy-tailed, power-law distributions rather than exponentials \cite{Mo2014}, a
phenomenon attributed to hydrodynamic entrapment. Such aging of the surface-attached state
justifies our implementation of a time-dependent reset probability in Phase $1$, where the particle
becomes increasingly ``persistent'' in its upstream navigation the longer it remains attached to the
boundary. As in the first case, both first-order and second-order DPT can be observed in the fluctuations 
of displacement of the particle depending on the aging strength. However, in contrast to the first case, we show 
that the aging logic of surface interactions shatters standard Gallavotti-Cohen symmetry and leads to a state of 
hibernation where the particle becomes trapped in a backward-biased state.

This paper is organized as follows: In Section~\ref{sec:2} we present a brief review of the theory of large deviations which 
deals with the probabilities of rare events with emphasizing on the systems with two sub-phases . In Section~\ref{sec:3} and 
Section~\ref{sec:4} we define and analyze two different examples. In Section~\ref{sec:5} we bring the concluding remarks.
\section{Large Deviation Theory: A Brief Review}\label{sec:2}
Let us consider the total displacement of a random walker $X_t$, sometimes called the current, integrated over $t$ time steps as a proper 
observable. The distribution of $X_t$ in the limit of large $t$ has a large deviation form \cite{To2009,To2025}:
\begin{equation} 
\label{P}
P(X_t/t=v) \approx e^{-t I(v)}.
\end{equation} 
This distribution is fully characterized, up to subleading corrections in $t$, by a rate function $I(v)$ defined as:
\begin{equation} 
I(v)=\lim_{t \to \infty} -\frac{1}{t} \ln P(X_t/t=v). 
\end{equation} 
Instead of considering $P(v)$ and $I(v)$, we can work with the generating function defied as:
\begin{equation} 
G(s,t)=\langle e^{s X_t} \rangle. 
\end{equation} 
The angular brackets denote an average over stochastic trajectories, started from some given initial distribution. The
generating function scales exponentially as: 
\begin{equation} 
G(s,t) \approx e^{t \Lambda(s)}
\end{equation} 
where the exponent which is called the SCGF is given by: 
\begin{equation}
\Lambda(s)=\lim_{t \to \infty} \frac{1}{t} \ln G(s,t). 
\end{equation} 
According to the G{\"a}rtner–Ellis theorem, given that $\Lambda(s)$ is differentiable, then $I(v)$ can be obtained as
the Legendre–Fenchel transform of the SCGF: 
\begin{equation} 
I(v)=\max_{s} \{ sv-\Lambda(s) \}.
\end{equation} 
The model we are studying in this paper consists of two sub-processes or phases called Phase $0$ and Phase $1$. 
Quite generally, we associate each phase with its own additive observable (current). 
Let $G_0(s, t)$ ($G_1(s, t)$ ) be the generating function for the current accumulated over $t$ steps in Phase $0$ ($1$). 
The total weight of a segment of length $t$ in Phase $0$ (denoted as $W_0$) and Phase $1$ (denoted as $W_1$) are
given by: 
\begin{align}
W_0(s, t) &= p_0(t) G_0(s, t), \label{Ws1} \\ 
W_1(s,t) &=  p_1(t) G_1(s, t)  \label{Ws2} 
\end{align}
in which $p_0(t)$ and $p_1(t)$ are generally discrete heterogeneous probability distributions. 
In this paper we assume that $p_0(t)$ and $p_1(t)$ are geometric distributions. This assumption means that 
the displacement is accumulated for $t-1$ steps, with the final transition step being current-neutral.
In order to calculate the generating function of the compound process, we adopt the approach used by 
Poland and Scheraga (PS) in studying the denaturation of the DNA \cite{PS19661,PS19662} and extended in \cite{Ha2017} for
studying the phase transitions in large deviations of reset processes. It turns out that it is easier to calculate 
the $z$-transform of the generating function of the compound process $G(s, z)$ given by : 
\begin{equation}
\widetilde{G}(s, z) = \sum_{t=1}^{\infty} {G}(s, t) z^{-t}= \frac{\widetilde{W_0}(s, z) + \widetilde{W_1}(s, z) + 2\widetilde{W_0}(s, z)\widetilde{W_1}(s, z)}
{1 - \widetilde{W_0}(s, z)\widetilde{W_1}(s, z)} 
\end{equation} 
in which $\widetilde{W_0}(s, z) $ and $\widetilde{W_1}(s, z) $ are the $z$-transform of  $W_0(s, t)$ and $W_1(s, t)$ respectively. 
The SCGF $\Lambda (s)$ is determined by the largest real root $z^*(s)$ of the denominator: 
\begin{equation} 
\label{TE} 
\widetilde{W_0}(s, z) \widetilde{W_1}(s,z) = 1. 
\end{equation} 
The SCGF is then $\Lambda (s) = \ln z^*(s)$. A DPT
occurs when $z^*(s)$ reaches the convergence boundary of either $\widetilde{W_0}(s, z)$ or 
$\widetilde{W_1}(s, z)$. This boundary is determined by the exponential growth of the sub-processes or the asymptotic
behavior of the age-dependent switching rate between them. It is worth mentioning that in \cite{Ha2017} the 
DPT comes from the time inhomogeneity of the sub-process (here Phase $0$ and Phase $1$). However, 
as we will see here, it results from time-dependent resets. 

As we mentioned, we consider two distinct examples. In following sections we analyze these two examples separately.
\section{The First Case: A Semi-Markovian Random Walk}\label{sec:3}
We start with writing the current generating function for a segment of $t$ consecutive steps in Phase $0$ as follows: 
\begin{equation} 
\label{w01}
W_0(s, t) = \left( p e^{s} + q e^{- s} \right)^{t -1}r(t)\prod_{i = 1}^{t - 1}{(1 - r(i))} 
\end{equation} 
in which $p$ ($q=1-p$) is the probability of hopping forward (backward) and $r(i)$ is the probability
of resetting to Phase $1$, or the wall-attached phase, at time $i$. The biasing field $s$ which counts the number of steps in both forward 
and backward directions, gives weight to the spatio-temporal trajectories so we can probe rare trajectories (corresponding to the rare 
values of the observable) in the so called $s$-ensemble. Following \cite{Ko2018} we assume that the reset probability $r(t)$ is given by: 
\begin{equation} 
r(t) =\frac{a}{b + t}\;\;  \text{for} \;\; t = 1,2,3,\ldots
\end{equation} 
with $0 < a < b + 1$ for the probability of resetting to be always less than $1$. The reader should note that in (\ref{w01}) the particle 
takes $t - 1$ consecutive steps and the last step at time $t$ is considered to be a reset to Phase $1$, as we expect from 
a geometric distribution. It is easy to check that the heterogeneous geometric distribution defined here is normalized \cite{Ma2007}: 
\begin{equation} 
\sum_{t = 1}^{\infty} p_0(t)=\sum_{t = 1}^{\infty}{r(t)}\prod_{i = 1}^{t - 1}\left( 1 - r(i) \right) = 1. 
\end{equation} 
For a segment of length $t$ in Phase $1$ the corresponding weight is defined as: 
\begin{equation} 
\label{w11}
W_1(s, t) = (1 - r(t))\prod_{i = 1}^{t -1}{r(i).} 
\end{equation} 
One can easily check that the normalization condition is fulfilled: 
\begin{equation} 
\sum_{t = 1}^{\infty} p_1(t)=\sum_{t =1}^{\infty}\left( 1 - r(t) \right)\prod_{i = 1}^{t - 1}{r(i)} = 1. 
\end{equation} 
The $z$-transform of (\ref{w01}) and (\ref{w11}) can be calculated and it turns out that they have a closed form: 
\begin{eqnarray} 
\widetilde{W_0}(s, z) &=& \frac{a {}_2F_1(1,1 - a + b,2 + b;\frac{p e^{s} + q e^{- s}}{z})}{z(1 + b)} , \\
\widetilde{W_1}(s, z)&=& \frac{e^{\frac{a}{z}}((1 - z)\Gamma(1 + b) - b\ \Gamma\left(
b,\frac{a}{z} \right) + z\ \Gamma(1 + b,\frac{a}{z}))}{z{(\frac{a}{z})}^{b}}
\end{eqnarray} 
in which $\Gamma(z)$ is the Gamma function, $\Gamma(a,z)$ is the incomplete Gamma function and ${}_2F_1(a,b,c;z)$ is the 
hypergeometric function. Before going into the detail of finding the SCGF $\Lambda(s)$, let us investigate the large-$t$ limit 
of (\ref{w01}) as it predicts the existence of DPTs in the fluctuations of the 
observable~\cite{Ha2017}. As we have already mentioned, a DPT occurs when for some value of $s$, the function $z^{*}(s)$ obtained from (\ref{TE}) reaches 
the convergence boundary point $z_{c}(s)$ of $\widetilde{W}_0(s,z)$ given by $z_{c}(s) = p e^{s} + q e^{- s}$. 
Note that $\widetilde{W}_1(z,s)$ is always convergent. For large $t$ one finds: 
\begin{equation} 
\label{asy1}
{W}_0(s,t)\sim\frac{{(p e^{s} + q e^{- s})}^{t}}{t^{a + 1}}. 
\end{equation} 
Comparing the above result with those of the PS model we realize that depending on the value of the aging strength $a$ there 
might be a DPT. From (\ref{asy1}) we find that for $0 < a \leq 1$ and $\forall b \geq 0$ two second-order DPTs
occur at $s_{c}^{(1)} = \ln\lbrack q/p\rbrack$ and $s_{c}^{(2)} = 0$; however, for $1 < a < 1 + b$ two first-order DPT occur at the same
critical points. These critical values $s_{c}^{(1,2)}$ can easily be calculated from $\ln z_{c}(s) = \ln\left( pe^{s} + qe^{- s} \right) = 0$
which clearly has two real roots. In summary the SCGF is given by $\Lambda(s) = \ln (pe^{s} + qe^{- s})$ for $- \infty < s \leq s_{c}^{(1)}$
and $s_{c}^{(2)} \leq s < \infty$ while it comes from (\ref{TE}) for $s_{c}^{(2)} \leq s \leq s_{c}^{(1)}$. Note that, without loss of generality, we
have assumed $p > q$. This assumption results in $s_{c}^{(1)} < 0$. 
\begin{figure}[t] 
\centering
\includegraphics[width=0.8\textwidth]{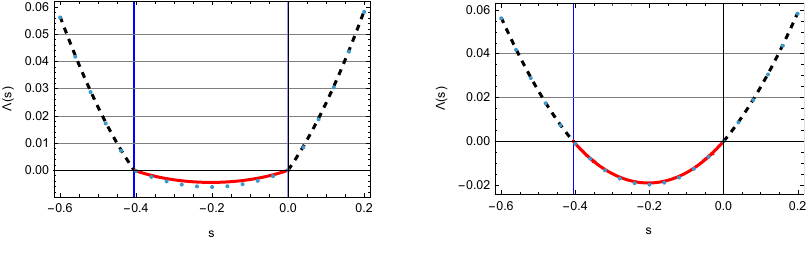} 
\caption{The SCGF $\Lambda (s)$ is plotted as a function the biasing field $s$. The black dashed line is
$\Lambda(s)=\ln (p e^s+q e^{-s})$. The red line is the solution of (\ref{TE}). The blue dotted line is the result of
the stochastic cloning simulation where the number of clones is $10^5$ and the run time is $500000$. 
The parameters are $p = 0.6,\ \ q = 0.4,\ \ b = 1.0,\ \ a = 1.5$ (left) and $a = 0.6$ (right). 
The vertical lines are the locations of the transition points $s_{c}^{(1)} = - 0.405$ and $s_{c}^{(2)}=0$.} 
\label{Fig1}
\end{figure} 
In Figure \ref{Fig1} we have plotted the SCGF $\Lambda (s)$ as a function of $s$ for two values of the aging strength $a$. 
It can be seen the SCGF is a smooth function of the biasing field $s$ and hence differentiable everywhere in $- \infty < s
< \infty$ (right in Figure \ref{Fig1}), while it is not differentiable at $s_{c}^{(1,2)}$ resulting in two linear
parts in the corresponding rate function $I(v)$ (left in Figure \ref{Fig1}). The blue dotted lines 
are the results of discrete time stochastic cloning simulations described in \cite{Ta2009}. The minor discrepancy from the exact results in the region $s_{c}^{(2)} \leq s \leq
s_{c}^{(1)}$ is due to the fact that considering a real large ensemble and $t \rightarrow \infty$ at the same time cannot be implemented in the
simulation; nevertheless, the stochastic cloning simulation predicts both the transition points and the overall behavior of the SCGF as $s
\rightarrow \pm \infty$ with a good degree of accuracy. In both cases, the fact that one of the singularity is always at $s = 0$ is very
special: it means the physical, unperturbed system is inherently critical, with no need for artificial biasing to reach the transition. The
distinction between first- and second-order DPT lies in whether the criticality involves coexistence and intermittency (at a first-order DPT) or
scale invariance and diverging susceptibility (at a second-order DPT). The significance of the transition at $s = 0$ can be explained in terms of
the dynamics of the system. $s = 0$ corresponds precisely to the unbiased probability measure on trajectories---the one that describes the
actual physical dynamics of the system as it evolves naturally under its own stochastic rules (no external tilting, conditioning, or
reweighting applied). This means that the critical point (where the singularity in the SCGF appears) is reached without any artificial
intervention. The phase transition, coexistence, or criticality is an inherent property of the system's parameter values (e.g., temperature,
density, interaction strength) in its standard, physical regime. In other words, the signatures of the transition are directly observable in
unbiased simulations, experiments, or real-world evolution. In contrast, when the transition occurs at some $s \neq 0$, the coexisting or
critical regimes would only dominate in a biased ensemble, which corresponds to conditioning the physical system on highly atypical
(exponentially rare) fluctuations. Observing those regimes directly would require either enormous observation times or sophisticated sampling
algorithms---making them physically inaccessible in practice without artificial aids. 
\subsection{Rate Function}
Finding an analytical expression for the SCGF in $s_{c}^{(2)} \leq s \leq s_{c}^{(1)}$ is generally a formidable task. However, as $a
\rightarrow 1 + b$ (where we have two first-order DPTs) one can see that the slope of the curve connecting the two critical
points becomes almost zero i.e. the two critical points $s_{c}^{(1,2)}$ are connected via a horizontal line. In this case, since the analytical
expression for the SCGF is known, the rate function $I(v)$ can be calculated exactly. The Legendre-Fenchel transform of the SCGF results in:
\begin{equation} 
\label{I}
I(v) = 
\begin{cases} 
f(v) & \text{for } -1 \le v \le -|p - q| \\ 
\ln\left( \frac{q}{p} \right)v & \text{for } -|p - q| \le v \le 0 \\ 
0 & \text{for } 0 \le v \le |p - q| \\ 
f(v) & \text{for } |p - q| \le v \le 1 
\end{cases} 
\end{equation} 
in which: 
\begin{equation}
f(v) = \frac{v}{2}\ln\left( \frac{q(1 + v)}{p(1 - v)} \right) - \ln\left( 2\sqrt{\frac{qp}{1 - v^{2}}} \right). 
\end{equation} 
The reader notes that $v \in \lbrack - 1,1\rbrack$ since the slope of the SCGF goes to $\pm 1$ as $s \to \pm \infty$ . 
As we mentioned before, the existence of a kink (non-analyticity in the first derivative) in the SCGF
results in linear parts in the rate function $I(v)$ and consequently the probability density function $P(v)$. 
This can be seen in (\ref{I}). Interestingly, since one of the kinks is located at the origin $s_{c}^{(2)} = 0$, we have two
linear parts in the rate function $I(v)$. The existence of a kink at the origin which results in a first-order DPT and coexistence of phases
has already been observed in different models including the kinetically constrained models of glass formers \cite{Ga2007}. 
\begin{figure}[t]
\centering 
\includegraphics[width=0.8\textwidth]{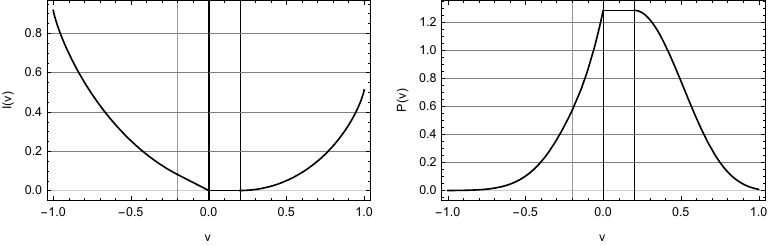} 
\caption{The rate function $I(v)$ (left) and the probability distribution function $P(v)$ (right)
are plotted as a function of $v$ for $p = 0.6$ and $t = 10$.} 
\label{Fig2}
\end{figure} 
In Figure \ref{Fig2} we have plotted both $I(v)$ given by (\ref{I}) and also properly normalized $P(v)$ given by (\ref{P}), for $p = 0.6$ and $t = 10$ as a 
function of $v$. As can be seen the rate function consists of four parts divided by vertical lines. The location of 
each vertical line is given in (\ref{I}). The flat part in $P(v)$ (or equivalently $I(v)$) is the direct signature of phase coexistence in the 
unbiased ($s = 0$) ensemble. It reflects the fact that, at the transition, the two dynamical phases (normal running and wall-detached) have 
equal statistical weight, and any mixture of them---corresponding to different fractions of trajectory time spent in each phase---costs 
no additional large deviation price. The system freely explores all possible macroscopic lever-rule combinations in space-time, leading 
to a uniform distribution over intermediate $v$. As in the equilibrium case, the coexistence does not exist at the second-order DPT point. 
However, since the analytical expression for the SCGF is not available for $s_{c}^{(2)} \leq s \leq s_{c}^{(1)}$ we have not been able to 
calculate the rate function in the case $0 < a \leq 1,\ \forall b \geq 0$.
\subsection{Mean Current at $s=0$}
To characterize the transport properties of the self-propelled particle at the unbiased physical point $s=0$, we employ the renewal reward theory \cite{Vl2014}. 
We define a complete cycle of the process as the combination of one normal running phase followed by one wall-attached phase. According to 
renewal theory, the long-term mean current $\langle v \rangle$ is given by the ratio of the expected displacement (reward) in one cycle 
to the expected duration of that cycle:
\begin{equation} 
\langle v \rangle = \frac{\mathbb{E}[X_{\text{0}}] + \mathbb{E}[X_{\text{1}}]}{\mathbb{E}[T_{\text{0}}] +
\mathbb{E}[T_{\text{1}}]} 
\end{equation} 
where $X$ denotes the displacement and $T$ denotes the sojourn time. In the wall-attached phase,
the particle is stationary, implying $\mathbb{E}[X_{\text{1}}] = 0$. In the normal running phase, we must account for the fact that the final step
of the sojourn corresponds to a reset event which is current-neutral. Consequently, the particle accumulates displacement only over
$T_{\text{0}} - 1$ steps. With a mean step velocity of $p - q$, the expected displacement in the running phase is
$\mathbb{E}[X_{\text{0}}] = (p - q) (\mathbb{E}[T_{\text{0}}] - 1)$. Thus, the expression for the mean current simplifies to:
\begin{equation} 
\langle v \rangle = (p - q) \frac{\mathbb{E}[T_{\text{0}}] - 1}{\mathbb{E}[T_{\text{0}}] + \mathbb{E}[T_{\text{1}}]}
\end{equation} 
The mean sojourn time in the normal running phase is determined by the survival probability $P_{\text{surv}}^{(0)}(t) =
\prod_{i=1}^{t-1} (1 - r(i))$, which for large $t$ decays as a power-law $t^{-a}$. Summing over the discrete time steps, we find the mean
sojourn time: 
\begin{equation} 
\label{ET0}
\mathbb{E}[T_{\text{0}}] = \sum_{t=1}^{\infty} \frac{\Gamma(b-a+t) \Gamma(b+1)}{\Gamma(b-a+1) \Gamma(b+t)} =
\begin{cases} 
\frac{b}{a-1} & 1 < a < 1+b \\ \infty & 0 < a \le 1 
\end{cases} .
\end{equation} 
The divergence of $\mathbb{E}[T_{\text{0}}]$ at $a=1$
marks the transition to the unbound regime where the particle effectively escapes the reset mechanism. Conversely, in the wall-attached phase,
the survival logic is inverted such that $P_{\text{surv}}^{(1)}(t) = \prod_{i=1}^{t-1} r(i)$. Due to the $1/\Gamma(t+b)$ scaling of this
product, the phase is characterized by a short-tailed distribution with a mean sojourn time that is always finite: 
\begin{equation}
\label{ET1}
\mathbb{E}[T_{\text{1}}] = \Gamma(b+1) \sum_{t=1}^{\infty} \frac{a^{t-1}}{\Gamma(b+t)} = {}_1F_1(1; b+1; a) 
\end{equation} 
where ${}_1F_1$ is the confluent hypergeometric function of the first kind. It is clear that the mean sojourn times depend on $a$ and $b$. 
Substituting these durations into the renewal formula for the $a > 1$ regime, we obtain: 
\begin{equation} 
\langle v \rangle = (p - q) \frac{b - a + 1}{b + (a-1) {}_1F_1(1; b+1; a)}. 
\end{equation} 
Combining both regimes, the steady-state current is:
\begin{equation} 
\langle v \rangle = 
\begin{cases} 
(p - q) \frac{b - a + 1}{b + (a-1) {}_1F_1(1; b+1; a)} & a > 1 \\ p - q & a \le 1
\end{cases} .
\end{equation} 
This result reveals a second-order phase transition in the parameter space of the model. 
\begin{figure}[t] 
\centering 
\includegraphics[width=0.8\textwidth]{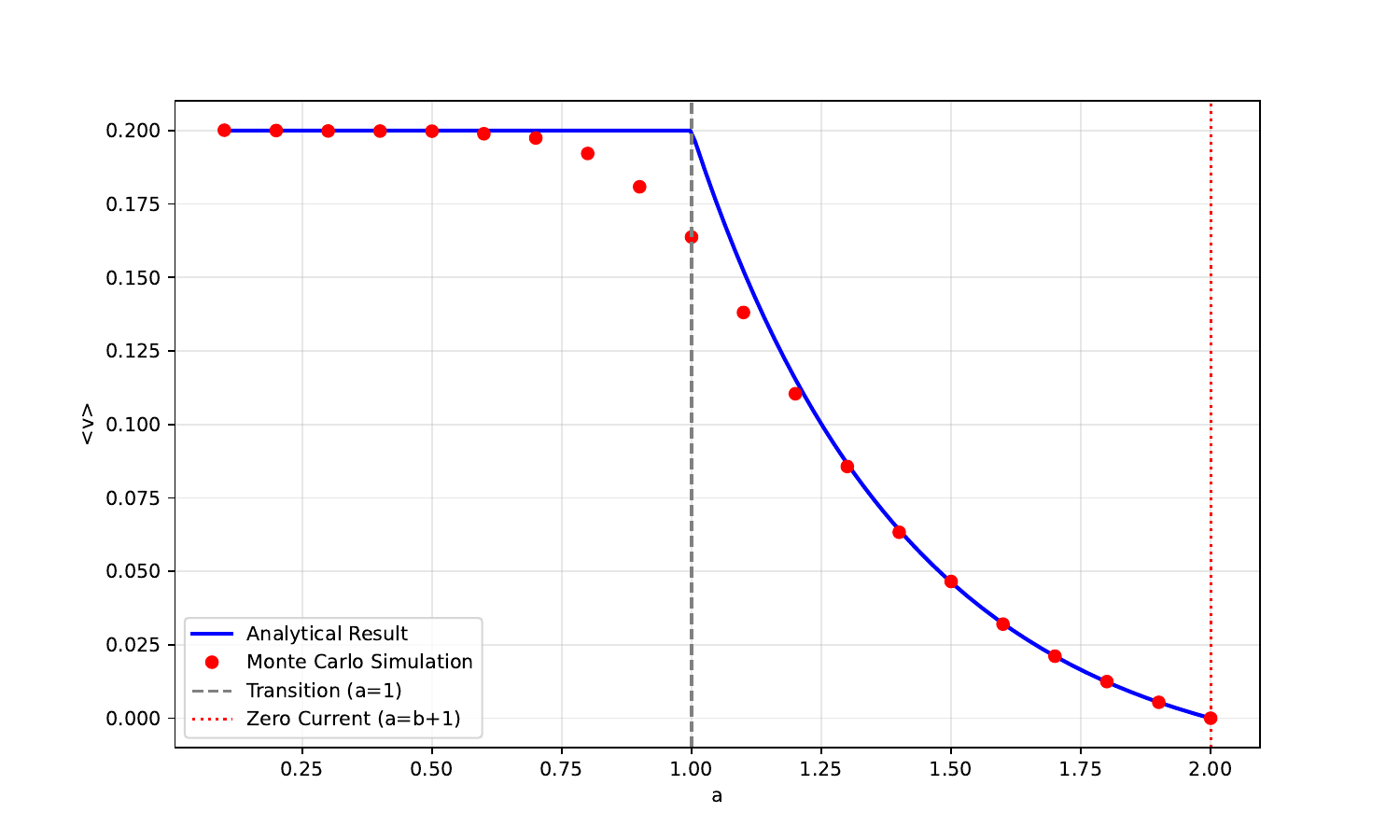} 
\caption{Mean current $\langle v \rangle$ at $s=0$ as a function of the aging strength $a$ for $p=0.6$ and $b=1.0$. The dashed vertical line at $a=1$ 
indicates the transition between the unbound regime (where the current is constant at $p-q$) and the bound regime (where resets reduce the current). 
This marks a continuous (second-order) transition in the parameter space. The dotted line is obtained from Monte Carlo simulations averaging over 
$60$ samples. The total time consists of $10^6$ steps.} 
\label{Fig3} 
\end{figure} 
For $a \le 1$, the particle is in the \textit{unbound regime}; despite the neutral switching step, the infinite mean duration of the running 
phase ensures the current converges to the full biased-walk velocity $p - q$. At $a = 1$, the system undergoes a continuous phase transition. 
For $a > 1$, the particle enters the \textit{bound regime}, where resets enforce a finite running time, causing the mean current to decrease 
monotonically with $a$. As shown in Figure \ref{Fig3}, while the current is continuous at $a=1$, the functional shift at this point 
represents a second-order transition in the parameter space. The Monte Carlo simulation results are shown as a red dotted line in Figure \ref{Fig3}.
The numerical discrepancy observed at $a=1$ arises from finite-time effects, as the analytical mean sojourn time $\mathbb{E}[T_{\text{0}}]$ 
diverges at this critical value. In Monte Carlo simulations, the finite total integration time naturally truncates the heavy power-law tails of the 
running phase, leading to a numerical estimate that converges slowly toward the theoretical infinite-time limit. 
\subsection{Conditional Mean Sojourn Time in Phase $1$} 
\begin{figure} 
\centering 
\includegraphics[width=0.8\textwidth]{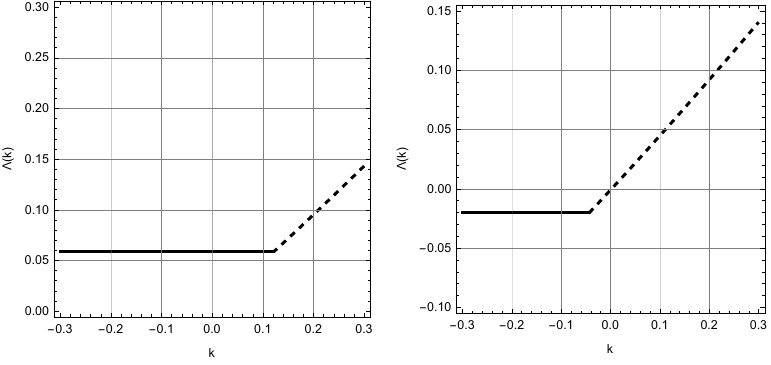} 
\caption{The
plot of the SCGF $\Lambda(k)$ for two values of the biasing field $s$: $s = + 0.2$ (left) and $s$=-0.2 (right). See the text for more detail.}
\label{Fig4}
\end{figure}

Finally, let us investigate the mean sojourn time of the particle in Phase $1$ conditioned on a fixed value of the velocity. 
According to the formalism explained in Section \ref{sec:2}, here we require two biasing fields that count the displacement 
in the normal running phase and the time steps spent in the wall-attached phase simultaneously. To this end, we introduce a new 
biasing field $k$ to count the time steps in the inactive phase and rewrite (\ref{w11}) as follows: 
\begin{equation} 
W_{1}(s,k,t) = e^{(t - 1)k}(1 - r(t))\prod_{i = 1}^{t - 1}{r(i)}. 
\end{equation} 
Its $z$-transform takes the following form: 
\begin{equation} 
\widetilde{W}_1(s,k,z) = \frac{ae^{\frac{ae^{k}}{z}}\left( \left( e^{k} - z \right)\Gamma(1 + b) - b\ e^{k}\Gamma\left(
b,\frac{ae^{k}}{z} \right) + z\ \Gamma\left( 1 + b,\frac{ae^{k}}{z} \right) \right)}{z^{2}\left( \frac{ae^{k}}{z} \right)^{b + 1}}.
\end{equation} 
Note that $\widetilde{W}_0$ does not change as we do not want to count the time steps in the running phase. Fixing $s = s^{*}$ and
solving the equation: 
\begin{equation}
\widetilde{W}_0(s^{*},k,z)\widetilde{W}_1(s^{*},k,z) = 1
\end{equation} 
for the largest root $z^{*}$ as a function of $k$ and then calculating $\Lambda (k) = \ln z^{*}$ gives the SCGF of the process given that the mean 
velocity is fixed. We remind the reader that the slope of the SCGF gives the mean value of the observable in the tilted ensemble. Therefore, fixing 
$s$ is equivalent to working in a tilted ensemble where the mean velocity is prescribed. Let us choose $a$ and $b$ so that $1 < a < 1 + b$. 
As we mentioned above, we have two first-order DPTs in this case. In Figure \ref{Fig4} we have plotted $\Lambda (k)$ for $s = + 0.2$ (left) 
and $s = - 0.2$ (right) for $a = 1.8$, $b = 1$ and $p = 0.6$. For $s = +0.2$ we are in the running phase. In this case the particle never resets to the 
Phase $1$; therefore, we expect that the typical mean sojourn time in Phase $1$ to be zero. The slope of $\Lambda (k)$ at $k=0$ is zero as it is seen in 
Figure \ref{Fig4} (left). In contrast, for $s = - 0.2$ the particle is in the mixed phase and as it can be seen in Figure \ref{Fig4} (right) 
the slope of the SCGF at $k=0$ is positive. As $k \to \infty$, we expect that the slope of $\Lambda (k)$ approaches to $1$ at $s=\pm 2$. 
This completes the characterization of the model in the first case.
\section{The Second case: A Particle in Opposing Flows}\label{sec:4}
Let us consider a biased random walker moving on a one-dimensional chain in discrete
time. The jump probability to the right is $p$ and to the left is $q=1-p$. As in the previous example,
we assume that the random walker resets between two sub-processes or phases: Phase $0$ and Phase
$1$. Phase $0$ is defined as a biased random walk with a preferred
direction ($p > q$ without loss of generality). The transition to Phase $1$ is governed by a
Markovian reset mechanism with a constant probability $r$. With constant reset probability $r$, the
survival probability $P_{\text{surv}}^{(0)}(t)$ is: 
\begin{equation} 
P_{\text{surv}}^{(0)}(t) = (1-r)^t = e^{t \ln(1-r)} \approx e^{-rt}.
\end{equation} 
The probability of staying in Phase $0$ drops exponentially fast toward zero. Upon
reset, the random walker enters Phase $1$. Compared to Phase $0$, the bias is now reversed
so that jump probability to the left is $p$ and to the right is $q$. The switching logic
incorporates age-dependence where the reset probability $r(t)$ is a decreasing function of the time
spent in Phase $1$ as $r(t)= a/(b+t)$ with $0 < a < b+1$ for $t = 1,2,3,\ldots$. Physically, the longer the random walker remains in the
Phase $1$, the more stable its attachment becomes. In Phase
$1$, the probability of staying at step $t$ is $(1 - \frac{a}{b+t}) = \frac{b+t-a}{b+t}$. The
survival function $P_{\text{surv}}^{(1)}(t)$ is the product: 
\begin{equation} 
P_{\text{surv}}^{(1)}(t) = 
\prod_{i=1}^{t} \left( \frac{b+i-a}{b+i} \right) = 
\frac{\Gamma(b-a+1+t)}{\Gamma(b-a+1)} \frac{\Gamma(b+1)}{\Gamma(b+1+t)}.
\end{equation} 
Using Stirling’s approximation for large $t$: 
\begin{equation} 
P_{\text{surv}}^{(1)}(t)  \sim t^{-a}
\end{equation} 
This creates a ``Heavy Tail,'' making long stays in Phase $1$ infinitely
more likely than in Phase $0$ for $a \le 1$. Note that if $a \le 1$, the mean
sojourn time in Phase $1$ diverges ($\mathbb{E}[T_1] \to \infty$), leading to Hibernation. 
As in the first case, the random walker has an internal
clock which resets every time it enters a new phase. Having the survival functions, one can easily
calculate the mean sojourn time in Phase $0$ and $1$. 

As in the first case we assume that every excursion in phase
$0$ definitely ends with a reset to phase $1$. This means that we are dealing with a simple
homogeneous geometric distribution $p_0(t)=r(1-r)^{t-1}$. The mean sojourn time in Phase $0$ is given2 by:
 \begin{equation} 
 \mathbb{E}[T_0] = \sum_{t=1}^{\infty} t \ p_0(t) = \sum_{t=1}^{\infty} t \ r(1-r)^{t-1} = \frac{1}{r} 
\end{equation}
Similarly, we assume that every excursion in phase $1$
definitely ends with a reset to phase $0$; however, here we are dealing with a heterogeneous
geometric distribution $p_1(t)=r(t)\prod_{i=1}^{t-1} (1-r(i))$. The mean sojourn time in phase $1$ is now given by:
\begin{equation} 
\mathbb{E}[T_1] = \sum_{t=1}^{\infty} t \ p_1(t) = \sum_{t=1}^{\infty} 
t \ r(t)\prod_{i=1}^{t-1} (1-r(i)) =
\begin{cases} 
\frac{b}{a-1} & 1 < a < 1+b \\ \infty & 0 < a \le 1 
\end{cases} 
\end{equation} 
which is identical to (\ref{ET0}) in the first case.

\begin{figure}[t] 
\centering
\includegraphics[width=0.7\textwidth]{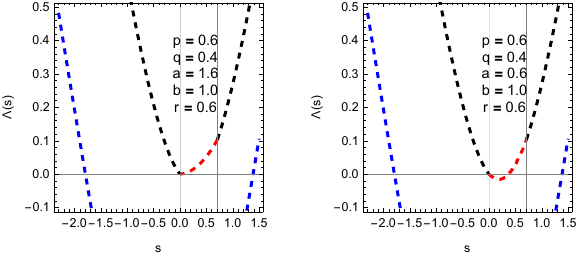} \caption{The plot of the SCGF $\Lambda (s)$ as a function of $s$. The vertical
lines are $s_{c}^{(1)}$ and $s_{c}^{(2)}$ given by (\ref{roots}). For $p>q$ the blue curve is always bellow
the red and the black curves; therefore, does not contribute in $\Lambda (s)$.} 
\label{Fig5} 
\end{figure}

We associate each phase with its own additive observable. Let $G_0(s, t)$
be the generating function for the current accumulated over $t$ steps in Phase $0$, and $G_1(s, t)$
for Phase $1$. The total weight of a segment of length $t$ in Phase $0$ and Phase
$1$ are given by (\ref{Ws1}) and (\ref{Ws2})
in which the homogeneous and the heterogeneous geometric distributions ($p_0(t)$  and $p_1(t)$) have been defined above. 
Displacement is accumulated for $t-1$ steps, with the final transition step being current-neutral; therefore, 
$G_0(s, t) = W_{\text{fwd}}(s)^{t-1}$, where $W_{\text{fwd}}(s) = p e^s + q e^{-s}$ and $G_1(s, t) =
W_{\text{bwd}}(s)^{t-1}$, where $W_{\text{bwd}}(s) = q e^s + p e^{-s}$. The $z$-transform of 
$W_0(s, t)$ for the phase $0$ is a rational function: 
\begin{equation} 
\widetilde{W}_0(s, z) = \sum_{t=1}^{\infty}
r(1-r)^{t-1} W_{\text{fwd}}(s)^{t-1} z^{-t} = \frac{r}{z - (1-r)W_{\text{fwd}}(s)} 
\end{equation}
The transform possesses a simple pole singularity at $z_0(s) = (1-r)W_{\text{fwd}}(s)$. The
$z$-transform of $V(s, t)$ for the phase $1$ is expressed via the hypergeometric function:
\begin{equation} 
\widetilde{W}_1(s, z) = \frac{a}{(b+1)z} 
{}_2F_1\left( b-a+1, 1; b+2; \frac{W_{\text{bwd}}(s)}{z} \right) 
\end{equation} 
The transform possesses a branch-cut singularity at $z_1(s) = W_{\text{bwd}}(s)$. 
Careful investigations show that in our model the SCGF comes from the solution of the transcendental
equation (\ref{TE}) for $s_{c}^{(1)} \le s \le s_{c}^{(2)}$ and $\Lambda (s)=\ln W_{\text{bwd}}(s)$
elsewhere. For $p > q$ the critical points $s_{c}^{(1)}$ and $s_{c}^{(2)}$ can be calculated from (\ref{TE})
at $z = W_{\text{bwd}}(s)$ that is: 
\begin{equation} \frac{r}{W_{\text{bwd}}(s_c) -
(1-r)W_{\text{fwd}}(s_c)} = W_{\text{bwd}}(s_c) 
\end{equation} 
This equation has two roots given by:
\begin{equation} 
\label{roots} 
s_{c}^{(1)}=0 \;\; \text{and}\;\; s_{c}^{(2)} = \frac{1}{2} \ln \left[
\frac{p(p - q + rq)}{q(q - p + rp)} \right].
\end{equation} 
The second root $s_{c}^{(2)}$ is called the re-entrant transition point as the system enters to a state 
of stagnant hibernation in Phase $1$ and exists as a finite positive value if $r > 1 - q/p$. In Figure
\ref{Fig5} we have plotted the SCGF as a function of $s$. The blue dashed line is $\Lambda
(s)=\ln W_{\text{fwd}}(s)$ which is always below the other curves. The black dashed line is $\Lambda
(s)=\ln W_{\text{bwd}}(s)$. Finally, the red dashed line is the solution of (\ref{TE}). The order of
the DPTs at $s_{c}^{(1,2)}$ is determined by the continuity of the slope
of the SCGF at the critical points i.e. if the mean current changes continuously or discontinuously
at the transition points. As in the first case, it turns out that they are both first-order or second-order depending on
the aging strength $a$: for $a\le 1$ the DPTs are both second-order while
for $a>1$ they are both first-order. 
Let us examine the validity of the Gallavotti-Cohen (GC) symmetry in the large-deviation analysis of this example. 
Given that the stagnant branches of the total SCGF are defined by the backward-biased aging Phase $1$, the 
natural reference symmetry for the system should be: 
\begin{equation} 
\Lambda(s) = \Lambda(E - s) 
\end{equation} 
where $E = \ln(p/q)$ represents the affinity magnitude. While the backward-evolution branches individually
satisfy this relation (with a center of symmetry at $s = E/2$), the introduction of the active
switching strategy shatters this global fluctuation symmetry. The numerical results in Figure~\ref{Fig5}
illustrate this breakdown vividly as the resetting valley (representing the switching strategy) emerges only for
positive fields $s > 0$ and does not possess a symmetric counterpart in the negative field region
relative to the $s = E/2$ axis. Physically, this symmetry breaking is a direct consequence of the
asymmetric memory structure; the persistent aging logic of the backward phase and the
compliant Markovian logic of the forward phase create a directional preference that the
statistical field cannot reconcile.
\subsection{A Note on Stability of the Switching Regime} 
The existence of the re-entrant critical point $s_{c}^{(2)}$ is conditional upon the reset probability $r$ from the 
forward-biased bulk phase (Phase $0$) to the wall-attached phase (Phase $1$). Analysis of the analytical
boundary condition reveals a critical threshold $r^* = 1 - q/p$. The value of $r$ relative to this
threshold defines two fundamentally different large-deviation behaviors. For $r > r^*$, the
switching is fragile, i.e., the resetting valley is finite, existing only within the range $s \in
[s_{c}^{(1)}, s_{c}^{(2)}]$. At high fluctuation fields, the ``cost'' of frequent resets from the forward phase
renders the switching strategy sub-optimal. The system undergoes a re-entrant transition, returning
to a state of stagnant hibernation in Phase $1$. In contrast, for $r \le r^*$, the switching is
robust. The denominator of the expression for $s_{c}^{(2)}$ becomes non-positive, and the second
critical point vanishes ($s_{c}^{(2)} \to +\infty$). In this regime, the forward Markovian phase is
sufficiently persistent that the energetic gain from forward sprints always outweighs the costs of
the switching cycle. Consequently, once the directional snap occurs at $s=0$, the particle remains in
the active switching regime for all positive fluctuations. This threshold represents a ``phase
stability transition'' in the parameter space. This suggests that rheotactic navigation is only robust 
against extreme fluctuations when the particle's bulk-to-wall transition probability is lower than its 
intrinsic normalized drift, defined by the ratio $(p-q)/p$.
\subsection{Slope of the SCGF at the critical points} 
We start with calculating the one-sided derivatives $\Lambda'(s)$ at $s_{c}^{(1)}=0$
for $a \le 1$. The left-sided derivative of the SCGF $s \to 0^-$ gives: 
\begin{equation}
\label{jleft1} 
\langle v_{\text{left}} \rangle = \lim_{s \to 0^-} \frac{d}{ds} \ln(q e^s + p e^{-s}) = q - p.
\end{equation} 
For the right-sided derivative $s \to 0^+$ we calculate the current by the
renewal reward theory. The long-time mean current is then given by the ratio of the expected
displacement in a single renewal cycle $\mathbb{E}[X_{\text{cyc}}]$ to the expected cycle duration
$\mathbb{E}[T_{\text{cyc}}]$. Noting that the final transition step of each segment is
current-neutral, we find: 
\begin{equation} 
\label{jright1} 
\langle v_{\text{right}} \rangle = \frac{(p-q)(\mathbb{E}[T_0] - 1) + (q-p)(\mathbb{E}[T_1] - 1)}{\mathbb{E}[T_0] + \mathbb{E}[T_1]} .
\end{equation} 
For $a \le 1$, $\mathbb{E}[T_1] \to \infty$ which results in
$\langle v_{\text{right}} \rangle = q - p$. This means that for the slope of the SCGF is continuous at $s_{c}^{(1)}$
for $a \le 1$. For $a > 1$ the formula (\ref{jleft1}) does not change; however, knowing
$\mathbb{E}[T_0] $ and $\mathbb{E}[T_1] $ the formula (\ref{jright1}) can be simplified and we find:
\begin{equation} 
\label{meanj} 
\langle v_{\text{right}} \rangle = (p-q) \frac{a - 1 - br}{a - 1 + br}.
\end{equation} 
This means that the slope of the SCGF is discontinuous at $s_{c}^{(1)}$.

In order to determine the order of the DPT at the re-entrant critical point
$s_{c}^{(2)}$, one can use  the implicit function theorem. The SCGF in the
switching valley is defined by the largest real root $z^*(s)$ of the master equation:
\begin{equation} 
F(s, z^*) = \widetilde{W}_0(s, z^*) \widetilde{W}_1(s, z^*) - 1 = 0 .
\end{equation} 
The derivative of the root with respect to the field $s$ is given by:
\begin{equation} 
\label{dzds}
\frac{dz^*}{ds} = - \frac{\partial_s F}{\partial_z F} = - \frac{(\partial_s
\widetilde{W}_0)\widetilde{W}_1+ \widetilde{W}_0(\partial_s \widetilde{W}_1)}{(\partial_z \widetilde{W}_0)\widetilde{W}_1+
\widetilde{W}_0(\partial_z \widetilde{W}_1)} .
\end{equation} 
The current in the valley is defined as $\langle v_{\text{val}}(s) \rangle = \frac{1}{z^*} \frac{dz^*}{ds}$. Calculating the partial derivatives 
and simplifying (\ref{dzds}) shows that for $a \le 1$ the valley current $\langle v_{\text{val}} \rangle$ matches the boundary current 
$\langle v_{\text{bwd}} \rangle$ and therefore  the transition is second-order.  
In contrast, for $a > 1$, $\langle v_{\text{val}} \rangle \neq \langle v_{\text{bwd}} \rangle$, resulting 
in a discontinuous jump in the current (a kink in the SCGF). 
Apart from this rigorous proof, the order of DPTs can be determined by examining the asymptotic
scaling of the sub-process generating functions, following the methodology established in \cite{Ha2017}. 
For the aging Phase $1$, the weight of a segment of duration $t$ scales as:
\begin{equation} 
W_1(s, t) \sim \frac{W_{\text{bwd}}(s)^t}{t^{a+1}} 
\end{equation} 
Mapping this to the PS framework, the exponent $c = a + 1$ dictates the convergence of the first derivative
of the $z$-transform at the branch-cut boundary $z = W_{\text{bwd}}(s)$. Since the switching valley
is bounded at both $s_{c,1}=0$ and $s_{c,2}$ by the same aging Phase $1$, the order of the
transitions is globally synchronized: For $a\le 1$ the exponent $c \le 2$ implies a diverging mean
sojourn time. The resulting singularity forces a tangential merge between the switching and stagnant
regimes at both critical points, characterizing a second-order DPT. For $a >
1$ the exponent $c > 2$ ensures a finite characteristic time scale. The first derivative of the
generating function remains finite at the boundary, resulting in a slope mismatch and a first-order
jump in the current at both critical points. This asymptotic approach confirms that the parameter
$a$ uniquely and universally determines the thermodynamic character of the resetting valley.
\begin{figure}[t] 
\centering \includegraphics[width=0.8\textwidth]{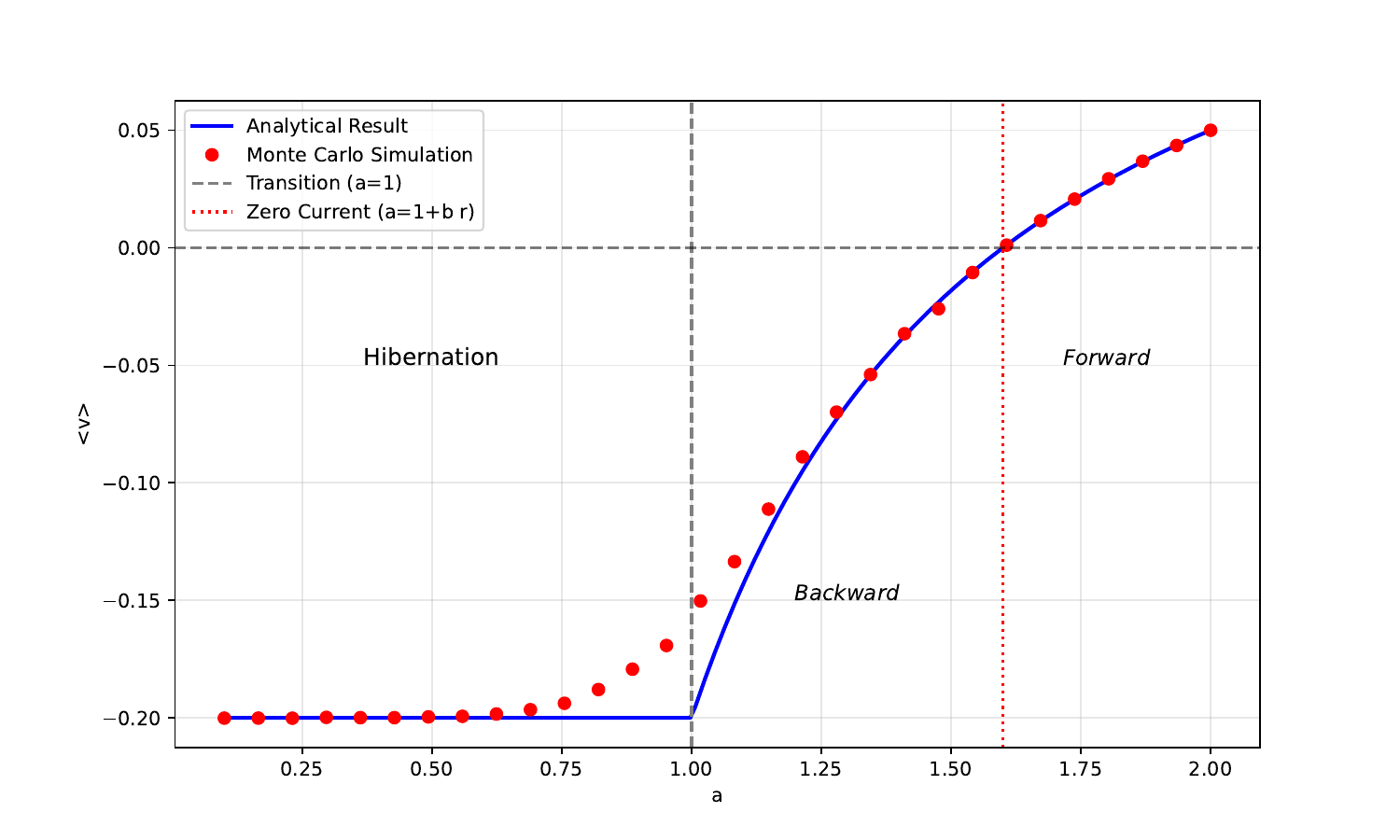} 
\caption{The mean current $\langle v \rangle$ as a function of the aging strength $a$ for $p=0.6$, $r=0.6$ and $b=1$. The point $a=1+ b \ r$ is where
the mean current becomes zero. The solid line is the exact analytical prediction while the dotted line is obtained from Monte Carlo simulations averaging over 
$60$ samples. The total time consists of $10^6$ steps.} 
\label{Fig6} 
\end{figure}
\subsection{Mean Current at $s=0$} 
The mean physical current $\langle v \rangle$ at the unbiased point $s_{c,1}=0$ has already been calculated. For $a \le 1$ 
we have found $\langle v \rangle = q - p$ while for $a > 1$ it is given by (\ref{meanj}). 
This expression identifies the hibernation limit $\langle v \rangle \to q-p$ as $a \to
1$ and the stalling point $\langle v \rangle= 0$ at $a = 1 + b\ r$, characterizing the steady-state directionality of
the particle. Monte Carlo simulations at $s=0$ confirm the $a$-space phase diagram. A
hibernation plateau exists for $a \le 1$ where $\langle v \rangle = q - p$. As shown in Figure
\ref{Fig6}, numerical results deviate from the theoretical plateau near $a=1$ due to
finite-time effects. Because the distribution is heavy-tailed, sampling rare excursions comparable
to the simulation length is limited, resulting in the rounding of the transition. A second-order transition in the parameter 
space is clear which occurs at $a=1$.
\section{Concluding Remarks} \label{sec:5}
This paper generalizes the approach first introduced in \cite{Ha2017} to study the effects of resetting 
and the occurrence of DPTs in two-state stochastic systems with time-heterogeneous dynamics, from  
large deviations viewpoint. We showed that time-heterogeneous resetting can likewise induce DPTs in 
such systems. This was illustrated through two detailed examples: a semi-Markovian random walk 
and a minimal model of rheotaxis. Notably, we observed that both continuous and discrete DPTs can 
occur at zero biasing field, depending on the aging strength. This finding highlights that time-dependent 
resetting alone—without an external bias—is sufficient to radically alter fluctuation behavior. 
The present work can be extended to stochastic dynamical systems with more than two internal states 
using the framework introduced in~\cite{Ei2010}.
 
\end{document}